\documentclass[prd,twocolumn,showpacs,superscriptaddress,nofootinbib,floatfix,showkeys,10pt]{revtex4-2}

\usepackage{graphicx}
\usepackage{amsmath}
\usepackage{bm}
\usepackage{yhmath}
\usepackage{mathtools}
\usepackage{wasysym}
\usepackage[colorlinks,citecolor=blue,urlcolor=blue,linkcolor=blue]
{hyperref}
\usepackage{subfigure}
\usepackage{color}
\usepackage{cases}
\usepackage{subfigure}
\usepackage{times}
\usepackage{dcolumn,booktabs,bm}
\usepackage{slashed}
\usepackage{amsfonts,amssymb,stmaryrd,latexsym,amsmath}
\usepackage{
textcomp}
\usepackage{multirow}
\usepackage{cancel}
\usepackage{array}
\usepackage{orcidlink}

\newcommand{\clabel}[2][]{#2}
\newcommand{\change}[1]{#1}

\def
\SU3{{\text{SU(3)}_{\rm F}}}

\def \Tcc{{T_{
cc}(3875)^+}}

\allowdisplaybreaks

\begin{document}


\title{Tetraquark bound states in constituent quark models: benchmark test calculations}

\author{Lu Meng\,\orcidlink{0000-0001-9791-7138}}\email{lu.meng@rub.de}
\affiliation{Institut f\"ur Theoretische Physik II, Ruhr-Universit\"at Bochum,  D-44780 Bochum, Germany }

\author{Yan-Ke Chen\,\orcidlink{
0000-0002-9984-163X}}\email{chenyanke@stu.pku.edu.cn}
\affiliation{School of Physics, Peking University, Beijing 100871, China}

\author{Yao Ma\,\orcidlink{0000-0002-5868-1166}}\email{yaoma@pku.edu.cn}
\affiliation{School of Physics and Center of High Energy Physics, Peking University, Beijing 100871, China}

\author{Shi-Lin Zhu\,\orcidlink{0000-0002-4055-6906}}\email{zhusl@pku.edu.cn}
\affiliation{School of Physics and Center of High Energy Physics, Peking University, Beijing 100871, China}


\begin{abstract}

We investigate the tetraquark bound states that are manifestly exotic using three distinct few-body methods: Gaussian Expansion Method (GEM), Resonating Group Method (RGM), and Diffusion Monte Carlo (DMC). We refer to manifestly exotic states that do not involve a mixture with the conventional mesons through the creation and annihilation of $n\bar{n}$, where $n=u, d$. Our calculations are conducted with two types of quark models: the pure constituent quark model featuring one-gluon-exchange interactions and confinement interactions, and the chiral constituent quark model, supplemented by extra one-boson-exchange interactions. This study represents a comprehensive benchmark test of various few-body methods and quark models. Our findings reveal the superiority of GEM over  RGM and DMC methods based on present implements for the tetraquark bound states. Additionally, we observe a tendency for the chiral quark model to overestimate the binding energies. We systematically explore the fully, triply, doubly, and singly heavy tetraquark states with $J^P=0^+,1^+,2^+$, encompassing over 150 states in total. We successfully identify several bound states, including $[cc\bar{n}\bar{n}]_{J^{P}=1^{+}}^{I=0}$, $[bb\bar{n}\bar{n}]_{J^{P}=1^{+}}^{I=0}$, $[bc\bar{n}\bar{n}]_{J^{P}=0^{+},1^{+},2^{+}}^{I=0}$, $[bs\bar{n}\bar{n}]_{J^{P}=0^{+},1^{+}}^{I=0}$, $[cs\bar{n}\bar{n}]_{J^{P}=0^{+}}^{I=0}$, and $[bb\bar{n}\bar{s}]_{J^{P}=1^{+}}$, all found to be bound states below the dimeson thresholds.

\end{abstract}

\maketitle

\thispagestyle{empty}

\section{Introduction}

Since the discovery of $X(3872)$~\cite{Belle:2003nnu}, many heavy-quarkonium-like states have been observed in experiments. These states are challenging to be accommodated within the quark-antiquark meson spectrum predicted by quark models, as exemplified in~\cite{Godfrey:1985xj, Barnes:2005pb}. They are considered as candidates of the tetraquark states (for recent reviews, see~\cite{Lebed:2016hpi, Chen:2016qju,Guo:2017jvc, Brambilla:2019esw, Liu:2019zoy, Chen:2022asf, Meng:2022ozq}). However, apart from states with the exotic quantum numbers~\cite{Adachi:2011mks, BESIII:2013qmu, Xiao:2013iha}, most of the heavy-quarkonium-like states may be a mixture of the tetraquark states and quark-antiquark states, influenced by the unquenched dynamics such as the creation and annihilation of the light $q\bar{q}$ pairs ($q=u,d,s$)~\cite{Eichten:2004uh, Barnes:2007xu, Lu:2016mbb}.

In the past three years, a series of exotic hadron states composed of at least four (anti)quarks have been observed. The LHCb collaboration first discovered the $X(6900)$ with a quark composition of $cc\bar{c}\bar{c}$~\cite{LHCb:2020bwg}. Subsequently, the CMS \cite{CMS:2023owd} and ATLAS~\cite{ATLAS:2023bft} collaborations confirmed the existence of the $X(6900)$ state and reported additional candidates for the fully charmed tetraquark states. In 2020, the BESIII collaboration reported the $Z_{cs}(3985)$ state with the minimal quark content $c\bar{c}\bar{s}u$ in the recoil-mass spectra of $K^+$ in the process $e^+e^-\to K^+ (D_s^-D^{*0}+D^{*-}_sD^0)$~\cite{BESIII:2020qkh}. Later, the LHCb also reported the state $Z_{cs}(4000)$ with the same quark contents but with a slightly higher mass and larger width. The LHCb collaboration also reported spin-0 and spin-1 states in the invariant mass spectrum of $D^-K^+$ in the decays $B^+\to D^+D^-K^+$\cite{LHCb:2020bls,LHCb:2020pxc}. These states were named as $T_{cs0}(2900)^0$ and $T_{cs1}(2900)^0$ according to the new naming convention \cite{Gershon:2022xnn}. The two states are candidates of the $cs\bar{u}\bar{d}$ tetraquarks. In addition, two charmed-strange tetraquark states, $T_{c\bar{s}0}(2900)^{++}$ and $T_{c\bar{s}0}(2900)^{0}$, with the minimal quark compositions of $c\bar{s}u\bar{d}$ and $c\bar{s}\bar{u}d$, were observed by LHCb~\cite{LHCb:2022lzp,LHCb:2022sfr}. The former is the first doubly charged tetraquark state observed in experiments, and the latter is likely its neutral partner. Furthermore, the doubly heavy tetraquark states, anticipated for about forty years~\cite{Ader:1981db,Zouzou:1986qh,Carlson:1987hh,Li:2012ss}, were also discovered. The $T_{cc}(3875)^+$ state, composed of $cc \bar{u}\bar{d}$, was observed by the LHCb collaboration~\cite{LHCb:2021auc,LHCb:2021vvq}. All the aforementioned states consist of four (anti)quarks if we neglect the unquenched effect of the heavy quark-antiquark pairs. Undoubtedly, we are rapidly entering the era of the ``genuine" multiquark states. These multiquark states have ignited heated theoretical discussions~\cite{Wang:2017uld,Meng:2020knc,Wang:2020htx,Meng:2021rdg,Meng:2021jnw,Wang:2021kfv,Du:2021zzh,Zhuang:2021pci,He:2022rta,Shi:2022slq,Niu:2022jqp,Wang:2022jop,Du:2023hlu,Wang:2023hpp,Hua:2023zpa,Wang:2023iaz}.

\begin{figure}[htp]
	\centering 
   \includegraphics[width=0.45\textwidth]{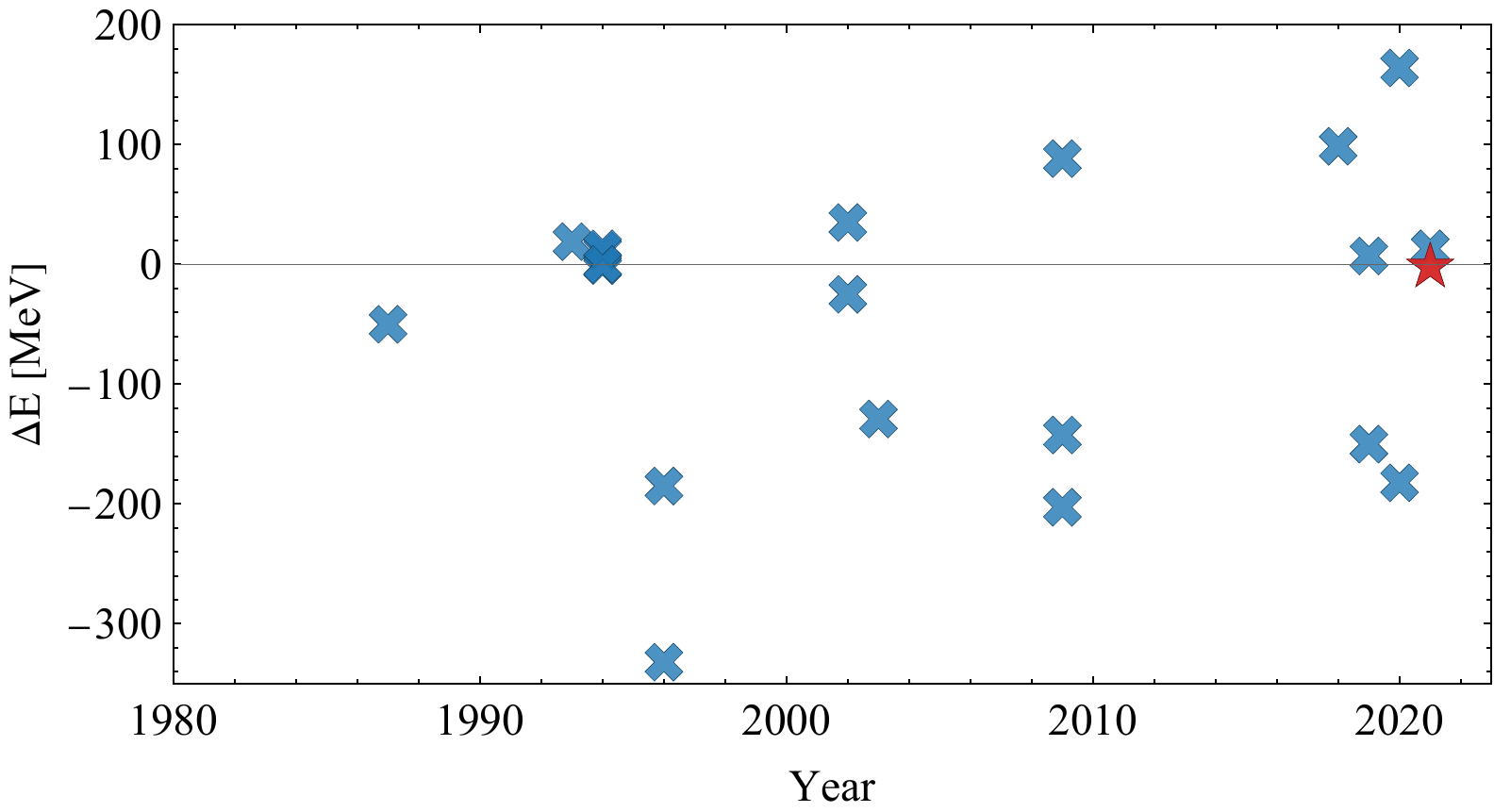}
	\caption{Theoretical predictions for the masses relative to the $DD^*$ threshold of the $c c \bar{n} \bar{n}$ tetraquark state with $I(J^P)=0(1^+)$~\cite{Carlson:1987hh,Silvestre-Brac:1993zem,Semay:1994ht,Pepin:1996id,Gelman:2002wf,Vijande:2003ki,Yang:2009zzp,Park:2018wjk,Maiani:2019lpu,Lu:2020rog,Tan:2020ldi,Noh:2021lqs,Yang:2019itm}, where theoretical uncertainties have been neglected. The red pentagram and blue cross correspond to the experimental result and theoretical predictions, respectively.}\label{fig:timeline}
\end{figure}

Historically, various quark models have made various predictions regarding tetraquark states. In this study, we focus exclusively on the nonrelativistic quark potential models. We do not consider quark models that parameterize the matrix elements without considering the spatial wave function~\cite{Cui:2006mp,Luo:2017eub,Eichten:2017ffp,Karliner:2017qjm,Cheng:2020wxa}, or incorporate relativistic effects~\cite{Ebert:2007rn,Lu:2020rog,Faustov:2021hjs} in our analysis. In Fig.~\ref{fig:timeline}, we present the case of the $c c \bar{n} \bar{n}$ state ($n=u,d$) with $I(J^P)=0(1^+)$ as an example to illustrate the predictions of various quark models. It is evident that these results exhibit significant divergence. Some calculations anticipate deeply bound states located below the $DD^*$ threshold with binding energies up to 300 MeV, while others suggest loosely bound states. Additionally, some calculations place the state above the $DD^*$ threshold, rendering it unstable. These significant variations can be attributed to differences in the potential models and the methodologies to solve the few-body problems.

In this study, we conduct benchmark calculations aimed at assessing the performance of various few-body methods to analyze the tetraquark bound states. Similar benchmark tests have previously been carried out for the four-nucleon systems~\cite{Kamada:2001tv}, in which seven different few-body methods were cross-referenced to compute the four-nucleon systems employing the same nucleon-nucleon (NN) interactions. For the tetraquark states, we will employ three distinct few-body methods, namely the Gaussian expansion method (GEM)\cite{Hiyama:2003cu}, the resonating group method (RGM)\cite{Entem:2000mq}, and the diffusion Monte Carlo method (DMC)~\cite{Gordillo:2020sgc,Gordillo:2021bra,Ma:2022vqf, Ma:2023int}. We check their consistency by comparing their results for the tetraquark bound states using the same quark potential models.

Another objective of this study is to explore the distinctions between different quark potential models for the tetraquark states. Unlike the nucleon systems where the NN scattering phase shifts constrain the interaction to several high-precision nucleon forces see~\cite{Stoks:1994wp,Machleidt:2000ge,Epelbaum:2019kcf}, the interactions among quarks exhibit greater variability. Basically, there are two types of constituent quark models: those featuring one-gluon-exchange (OGE) interaction combined with the confinement interaction, and those encompassing both of these interactions alongside an additional one-boson-exchange (OBE) interaction. The latter category is referred to as the chiral constituent quark models ($\chi$CQM), owing to the inclusion of the pseudoscalar meson exchange interactions arising from the spontaneous breaking of chiral symmetry. In this work, we designate the former quark models without OBE as the pure constituent quark models (PCQM). The debate over which type of quark models is superior has persisted for many years, yet without definitive conclusions. For instance, it has been argued that the two types of quark models yield qualitatively consistent baryon-baryon scattering results~\cite{Wang:2002ha}. In this study, we select the AL1 and AP1 models~\cite{Semay:1994ht,Silvestre-Brac:1996myf} as the representative examples of PCQM, while we employ a quark model proposed by the group at Salamanca University~\cite{Vijande:2004he,Segovia:2011dg} as an example of a chiral quark model.

After testing the reliability of the quark potential models and few-body methods, we proceed to predict the tetraquark states located below the strong decay threshold. We focus on the tetraquark states that have no admixture with the conventional mesons due to the creation and annihilation of the $n\bar{n}$ pairs, where $n=u,d$. It is important to note that we loosen the constraint and assume the unquenched effect of the $s\bar{s}$ pairs is suppressed. In our investigation, we examine systems of $QQ\bar{Q}\bar{Q}$, $QQ\bar{Q}\bar{q}$, $QQ\bar{q}\bar{q}$, and $Qq\bar{q}\bar{q}$ with $J^P=0^+, 1^+$, and $2^+$, encompassing a total of over 150 systems. Here, $Q=c,b$ and $q=u,d,s$.

The paper is organized as follows. In Sec.~\ref{sec:formalism}, we present an introduction to three different quark models and three distinct few-body methods. In Sec.~\ref{sec:num}, we provide a detailed exploration of the possible bound solutions, including the fully, triply, doubly, and singly tetraquark states. In this section, we offer a comprehensive comparison of the results obtained from different models and different few-body methods, alongside a direct comparison with the lattice QCD results. In Sec.~\ref{sec:sum}, we summarize our findings and assessment of various quark models and few-body methods. We also list the final candidates of the tetraquark bound states against strong decays.

\section{Formalism}~\label{sec:formalism}
\subsection{Constituent quark models}
In this work, we only focus on the nonrelativistic quark models with the Hamiltonian,
\begin{equation}
  H=\sum_i^4\left(m_i+\frac{\boldsymbol{p}_i^2}{2m_i}\right)-T_{CM}+\sum_{i<j}V_{ij}(r_{ij})\,,
\end{equation}
where the kinetic energy of the center of mass $T_{CM}$  is subtracted. We only consider the pairwise interaction. One can find the effect of other kinds of interactions in Refs.~\cite{Ma:2022vqf,Wang:2023igz}.

A minimal quark model consists of the OGE interaction and confinement interaction. In this work, we choose the quark models proposed in Refs.~\cite{Semay:1994ht,Silvestre-Brac:1996myf},
\begin{align}V_{ij}^{\text{AL1/AP1}}=-\frac{3}{16} & \lambda_{i}^{c}\cdot\lambda_{j}^{c}\Big(-\frac{\kappa}{r_{ij}}+\lambda r_{ij}^{p}-\Lambda \nonumber\\&
+\frac{2\pi\kappa'}{3m_{i}m_{j}}\frac{\exp(-r_{ij}^{2}/r_{0}^{2})}{\pi^{3/2}r_{0}^{3}}\boldsymbol{\sigma}_{i}\cdot\boldsymbol{\sigma}_{j}\Big),
\end{align}\label{eq:AL1AP1}
where the power coefficient is 1 for AL1 model and 2/3 for AP1 model. The $\lambda^c_i$ is the Gell-Mann matrix for the SU(3) color symmetry, and $\sigma_i$ is the Pauli matrix in the spin space. $m_i$ is the quark mass. $\kappa$, $\kappa'$ and $\lambda$ are the coupling constants for the Coulomb interaction, Gaussian hyperfine interaction and confinement interaction, respectively. $r_0$ and $\Lambda$ are the typical scale of the hyperfine interaction and overall shift parameter, respectively. All the above parameters were determined by the meson and baryon spectra. One can find their specific values in Ref.~\cite{Semay:1994ht}. For the minimal quark model, one can find different choices of parameter sets in Refs.~\cite{Liu:2019zuc}

For the $\chi$CQM, we choose the quark model proposed by the group of University of Salamanca (SLM for short) as an example. The idea of this model can be traced back to the Ref.~\cite{Entem:2000mq} to depict the NN system. The specific form was set up in Ref.~\cite{Vijande:2004he} and the parameters were redetermined in Ref.~\cite{Segovia:2011dg} by fitting the meson spectrum. The interactions read,
\begin{align}    V_{ij}^{\text{SLM}}&=\lambda_{i}^{c}\cdot\lambda_{j}^{c}\Biggl[\frac{\alpha_{s}}{4}\left(\frac{1}{r_{ij}}-\frac{1}{6m_{i}m_{j}}\frac{e^{-r_{ij}/r_{0}}}{r_{0}^{2}r_{ij}}\bm{\sigma}_{i}\cdot\bm{\sigma}_{j}\right)\nonumber\\&
+\left(-a_{c}(1-e^{-\mu_{c}r_{ij}})+\Delta\right)\Biggr]+V_{ij}^{\text{OBE}},
\end{align}\label{eq:SLM}
where the $\lambda^c_i$ is the Gell-Mann matrix for the SU(3) color symmetry and $\sigma_i$ is the Pauli matrix in the spin space. $m_i$ is the quark mass. $\alpha_s$, $a_c$ are the coupling constants for the OGE interaction and confinement interaction, respectively. Here, the Yukawa-type hyperfine interaction is chosen with a typical scale $r_0$. $\Delta$ is the overall shift parameter. For the confinement potential, the color screening effect is included, which becomes a linear interaction at the short distance and a constant at the long distance. In addition to the OGE and confinement interaction, the pseudoscalar meson exchange is included considering the spontaneous breaking of the chiral symmetry~\cite{Manohar:1983md}. Meanwhile, the meson-exchange interaction is extended to the scalar-meson-exchange interaction to mimic the two-pion-exchange interaction. All the above OBE interactions read,
\begin{align}V_{ij}^{\mathrm{OBE}}= & \ V_{ij}^{\pi}\sum_{a=1}^{3}\left(\lambda_{i}^{a}\cdot\lambda_{j}^{a}\right)+V_{ij}^{K}\sum_{a=4}^{7}\left(\lambda_{i}^{a}\cdot\lambda_{j}^{a}\right)\nonumber\\
 & +V_{ij}^{\eta}\left[\cos\theta_{P}\left(\lambda_{i}^{8}\cdot\lambda_{j}^{8}\right)-\sin\theta_{P}\right]\nonumber\\
 & +V_{ij}^{\sigma}\,,
\end{align}\label{eq:Vobe}
with 
\begin{align}V_{ij}^{\chi}= & \frac{g_{\mathrm{ch}}^{2}}{4\pi}\frac{m_{\chi}^{2}}{12m_{i}m_{j}}\frac{\Lambda_{\chi}^{2}}{\Lambda_{\chi}^{2}-m_{\chi}^{2}}m_{\chi}\left(\boldsymbol{\sigma}_{i}\cdot\boldsymbol{\sigma}_{j}\right)\nonumber\\
 & \times\left[Y\left(m_{\chi}r_{ij}\right)-\frac{\Lambda_{\chi}^{3}}{m_{\chi}^{3}}Y\left(\Lambda_{\chi}r_{ij}\right)\right]\,,\ \chi=\pi,K,\eta\,\\
V_{ij}^{\sigma}= & -\frac{g_{\mathrm{ch}}^{2}}{4\pi}\frac{\Lambda_{\sigma}^{2}}{\Lambda_{\sigma}^{2}-m_{\sigma}^{2}}m_{\sigma}\left[Y\left(m_{\sigma}r_{ij}\right)-\frac{\Lambda_{\sigma}}{m_{\sigma}}Y\left(\Lambda_{\sigma}r_{ij}\right)\right]\,,\nonumber
\end{align}\label{eq:VOPE2}
where $\lambda^a$ is the Gell-Mann matrix in the SU(3) flavor symmetry. $Y(x)=e^{-x}/x$ is the Yukawa function. The coupling constant $g_{ch}$ is determined by the experimental $NN\pi$ vertex~\cite{Entem:2000mq}. $\theta_P$ is the mixing angle to introduce the physical $\eta$ rather than the one in the octet of the SU(3) symmetry limit. $m_\chi$ are the experimental masses for the $\pi$, $K$ and $\eta$ mesons. The $m_\sigma$ is determined via the PCAC relation $m^2_\sigma\sim m_\pi^2+4m^2_{u,d}$. The cutoff $\Lambda_\chi$ and $\Lambda_\sigma$ are determined by fitting the meson spectrum. In this work, we use the parameter values in Ref.~\cite{Segovia:2011dg}. It is worthwhile to mention that the vector-meson-exchange interactions are also incorporated in some chiral quark mdoels~\cite{Wang:2011rga,He:2023ucd}.

With the three different quark models, we present the theoretical ground meson spectra in Fig.~\ref{tab:meson}. One can see their numerical results agree with the experimental results well. For simplicity, we assume there is no mixing effect between $\eta(n\bar{n})$ with $I=0$ and $\eta(s\bar{s})$, which are irrelevant to our tetraquark bound state predictions.

\begin{table*}[htp]
    \centering
        \caption{Mass spectra of the ground state mesons from three different quark models (in units of MeV). The ``Exp." represents the experimental results~\cite{ParticleDataGroup:2022pth} as a comparison.}
    \label{tab:meson}
    \begin{tabular*}{\hsize}{@{}@{\extracolsep{\fill}}cccccccccccc@{}}
\hline \hline
$J^P=0^-$ & $\pi$ & $\eta(n\bar{n})$~\footnote{For simplicity, we assume there is no mixing effects between $\eta(n\bar{n})$ with $I=0$ and $\eta(s\bar{s})$, which are irrelevant to our tetraquark bound state predictions.} & $\eta(s\bar{s})$ & $K$ & $D$ & $D_{s}$ & $B$ & $B_{s}$ & $B_{c}$ & $\eta_{c}$ & $\eta_{b}$\tabularnewline
 \hline 
Exp. & 139.57 & 547.86 & 957.78 & 493.68 & 1869.7 & 1968.4 & 5279.3 & 5366.9 & 6274.5 & $\text{2983.9}$ & 9398.7\tabularnewline
AL1 & 138.16 & 138.16 & 713.00 & 490.92 & 1862.4 & 1962.5 & 5293.5 & 5361.0 & 6291.6 & 3005.3 & 9423.7\tabularnewline
AP1 & 138.95 & 138.95 & 700.9 & 498.22 & 1881.3 & 1954.8 & 5311.2 & 5355.6 & 6268.6 & 2982.4 & 9401.2\tabularnewline
SLM & 139.76 & 686.96 & 813.77 & 468.62 & 1896.1 & 1983.3 & 5274.8 & 5347.5 & 6274.5 & 2989.5 & 9451.2\tabularnewline
\hline 
$J^P=1^-$ & $\rho$ & $\omega$ & $\phi$ & $K^{*}$ & $D^{*}$ & $D_{s}^{*}$ & $B^{*}$ & $B_{s}^{*}$ & $B_{c}^{*}$ & $J/\psi$ & $\Upsilon$\tabularnewline
\hline 
Exp. & 775.26 & 782.66 & 1019.5 & 891.67 & 2010.3 & 2112.2 & 5324.7 & 5415.4 & 6328.9 & 3096.9 & 9460.3\tabularnewline
AL1 & 767.00 & 767.00 & 1020.8 & 903.55 & 2016.1 & 2102.0 & 5350.5 & 5417.5 & 6343.2 & 3101.3 & 9461.5\tabularnewline
AP1 & 770.12 & 770.12 & 1021.4 & 907.56 & 2033.1 & 2106.9 & 5367.3 & 5418.0 & 6337.8 & 3102.5 & 9460.6\tabularnewline
SLM & 773.03 & 692.7 & 1000.3 & 901.90 & 2017.7 & 2110.8 & 5316.8 & 5393.3 & 6328.9 & 3096.8 & 9502.0\tabularnewline
\hline\hline   
\end{tabular*}
\end{table*}

\subsection{Gaussian expansion method}\label{subsec:GEM}

The first few-body method used to solve the tetraquark systems is the Gaussian expansion method~\cite{Hiyama:2003cu}. Namely, we expand the spatial wave function of $\bm{r}$ using the following basis,
\begin{equation}
    \phi_{nlm}(\bm{r})=\sqrt{\frac{2^{l+5/2}}{\Gamma(l+\frac{3}{2})r_{n}^{3}}}\left(\frac{r}{r_{n}}\right)^{l}e^{-\frac{r^{2}}{r_{n}^{2}}}Y_{lm}(\hat{r})\label{eq:gauss}
\end{equation}
where the $r_n$ is taken in geometric progression, $r_n=r_0a^{n-1}$. $Y_{lm}$ is the spherical harmonics representing the angular momentum. The basis functions are not orthogonal but could be approximately complete if a large range of $r_n$ were taken. It has been proved that the choice of the basis can embed both long- and short-range correlations simultaneously~\cite{Hiyama:2003cu}. 

\begin{figure}[htp]
	\centering  \includegraphics[width=0.4\textwidth]{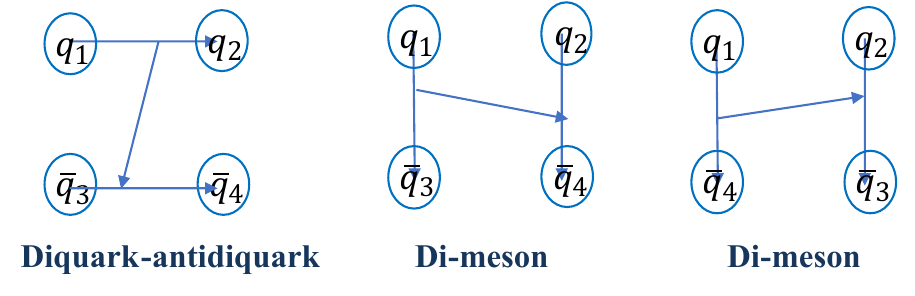}
	\caption{Jacobi coordinates used in the GEM of this work.}\label{fig:jacobi}
\end{figure}

For a four-body system omitting the motion of the center of mass, there are three independent coordinates. As depicted in Fig.\ref{fig:jacobi}, various sets of Jacobi coordinates can be chosen.  In principle, different choices of the set of Jacobi coordinate will give the same results if the basis functions are complete. One can choose either set of Jacobi coordinates and construct the basis with the total angular momentum $J$ by combining the spatial angular momenta regarding three coordinates and the spin wave functions. To make the basis function complete, the orbital excited basis functions should be incorporated. However, it takes great pains to handle the angular momentum in GEM, although the strategy has been invented~\cite{Hiyama:2003cu}. Alternatively, we only use the $l=0$ spatial wave functions in Eq.~\eqref{eq:gauss} but include different Jacobi coordinates to consider the different spatial correlations. In our calculation, we include three different sets of Jacobi coordinates as shown in Fig.~\ref{fig:jacobi}. For each coordinate, we choose six basis functions with $r_n$ in geometric progression. In order to make the basis functions more efficient, we choose different $r_0$ and $r_{max}$,
\begin{equation}
\begin{cases}
r_{0}=0.1\text{ fm},r_\text{max}=2\text{ fm} & q-q \text{ or } \bar{q}-\bar{q}\\
r_{0}=0.1\text{ fm},r_\text{max}=2\text{ fm} & (qq)-(\bar{q}\bar{q})\\
r_{0}=0.1\text{ fm},r_\text{max}=1\text{ fm} & q-\bar{q}\\
r_{0}=0.1\text{ fm},r_\text{max}=5\text{ fm} & (q\bar{q})-(q\bar{q})
\end{cases},
\end{equation}
where we take a large $r_\text{max}$ for the spatial wave functions between two $q\bar{q}$ clustering to depict the possible molecular solutions. In general, there are four extra $K$-type coordinates as mentioned in Refs.~\cite{Hiyama:2003cu,Meng:2020knc}. We have verified that the current selection without them already yields very precise results.

In addition to the spatial wave functions, we also have different options to construct the discrete wave function. For the color wave functions, one could use either of the following color wave functions,
\begin{eqnarray}
&\text{color-I: }\begin{cases}
[(q_{1}q_{2})_{\bar{3}}(\bar{q}_{3}\bar{q}_{4})_{3}]_{1}\\{}
[(q_{1}q_{2})_{6}(\bar{q}_{3}\bar{q}_{4})_{\bar{6}}]_{1}
\end{cases} \label{eq:colorI},\\
&\text{color-II: }\begin{cases}
[(q_{1}\bar{q}_{3})_{1}(q_{2}\bar{q}_{4})_{1}]_{1}\\{}
[(q_{1}\bar{q}_{4})_{1}(q_{2}\bar{q}_{3})_{1}]_{1}
\end{cases} \label{eq:colorII},\\
&\text{color-III: }\begin{cases}
[(q_{1}\bar{q}_{3})_{1}(q_{2}\bar{q}_{4})_{1}]_{1}\\{}
[(q_{1}\bar{q}_{3})_{8}(q_{2}\bar{q}_{4})_{8}]_{1}
\end{cases}.
\end{eqnarray}
The color-I is the diquark-antidiquark basis in the color space. Color-II represents the dimeson basis, where two basis are not orthogonal but complete. The color-III is orthogonal and complete. For the spin-wave functions, one can choose one of the following basis functions,
\begin{eqnarray}
  &&  \text{spin-I: }\begin{array}{c}
S_{12}=0,1;\;S_{34}=0,1\\
S_{12}\otimes S_{34}\to J
\end{array},\\
&&\text{spin-II: }\begin{array}{c}
S_{13}=0,1;\;S_{24}=0,1\\
S_{13}\otimes S_{24}\to J
\end{array},\\
&&\text{spin-III: }\begin{array}{c}
S_{14}=0,1;\;S_{23}=0,1\\
S_{14}\otimes S_{23}\to J\label{eq:spinIII}
\end{array},
\end{eqnarray}
where the S-wave orbital angular momentum is assumed. The spin-I is the diquark-antidiquark basis and spin-II and spin-III are two dimeson basis. Each of them is complete. In addition to above coupling modes in color and spin, one can also construct the discrete basis in a sequence of combining three (anti)quarks first and then with the forth (anti)quark. Similarly, one can find possible options of the complete flavor wave functions. 

\clabel[antism]{In our calculations, the wave functions of tetraquark states are expressed as the direct product of flavor wave function $\chi_f$, color-spin wave function $\psi_{cs}$, and spatial wave function $\psi$:
\begin{equation}
\Psi={\cal A}(\chi_{f}\otimes\psi_{cs}\otimes\psi),
\end{equation}
Here, ${\cal A}$ denotes the antisymmetrization operator, representing the exchange of identical quarks. For instance, in the case of $bb\bar n\bar n$ states, the antisymmetrization operator is defined as ${\cal A}_{12,34}=(1-P_{12})(1-P_{34})$, whereas for $bc\bar n\bar n$ states, it becomes ${\cal A}_{34}=(1-P_{34})$, where $P_{ij}$ permutes the $i$-th and $j$-th (anti)quarks. Our approach involves constructing basis wave functions with fixed quantum numbers, followed by the application of the antisymmetrization operator. It is important to note that antisymmetrization introduces additional constraints, potentially reducing the basis space. In other words, independent basis functions may become linearly dependent after antisymmetrization. To address this, we employ the algorithm outlined in Appendix~\ref{app:reduce} to automatically eliminate redundant bases in our calculations. }

In our calculation, we test five different choices of wave functions numerically,
\begin{eqnarray}
  \Psi_{A}	&=& {\cal A}\left(\chi_{f}\otimes\chi_{cs}^{\text{All}}\otimes\psi^{\text{All}}\right)\,,\nonumber\\
  \Psi_{B}	&=&{\cal A}\left(\chi_{f}\otimes\chi_{cs}^{\text{dimeson}}\otimes\psi^{\text{All}}\right)\,,\nonumber\\
\Psi_{C}	&=&{\cal A}\left(\chi_{f}\otimes\chi_{cs}^{\text{diquark}}\otimes\psi^{\text{All}}\right)\,,\nonumber\\
\Psi_{D}	&=&{\cal A}\left(\chi_{f}\otimes\chi_{cs}^{\text{All}}\otimes\psi^{\text{\text{dimeson}}}\right)\,,\nonumber\\
\Psi_{E}	&=&{\cal A}\left(\chi_{f}\otimes\chi_{cs}^{\text{All}}\otimes\psi^{\text{\text{diquark}}}\right)\,,\label{eq:GEMwave}
\end{eqnarray}
with
\begin{eqnarray}
    \chi_{cs}^{\text{diquark}}	&\equiv &\chi_{c}^{\text{I}}\otimes\chi_{s}^{\text{I}}\,,\nonumber\\
\chi_{cs}^{\text{dimeson}}	&\equiv&\left(\chi_{c}^{\text{II,1}}\otimes\chi_{s}^{\text{II}}\right)\oplus\left(\chi_{c}^{\text{II,2}}\otimes\chi_{s}^{\text{III}}\right)\,,\nonumber\\
\psi^{\text{All}}	&\equiv&\psi^{\text{\text{diquark}}}\oplus\psi^{\text{dimeson}}\,,\nonumber\\
\chi_{cs}^{\text{All}}	&\equiv&\chi_{cs}^{\text{\text{diquark}}}\oplus\chi_{cs}^{\text{dimeson}},
\end{eqnarray}
where $\chi_f$, $\chi_s$, $\chi_c$ and $\psi$ represent flavor, spin, color and spatial functions, respectively. Their superscript shows the specific choice of the wave functions in Eqs.~\eqref{eq:colorI}-\eqref{eq:spinIII} and Fig~\ref{fig:jacobi}. The $\chi_c^\text{II,1}$ and $\chi_c^\text{II,2}$ are the first and second color basis in Eq.~\eqref{eq:colorII}, respectively.  Among them, $\Psi_A$ is the most general basis. $\Psi_B$ and $\Psi_C$ are basis wave functions with general spatial wave function but dimeson and diquark-antidiquark discrete wave functions, respectively. The discrete wave functions of $\Psi_D$ and $\Psi_E$ are general but with dimeson and diquark-antidiquark spatial wave functions, respectively.

We use $[cc\bar{n}\bar{n}]_{J^{P}=1^{+}}^{I=0}$, $[bb\bar{n}\bar{n}]_{J^{P}=1^{+}}^{I=0}$ and  $[bc\bar{n}\bar{n}]_{J^{P}=2^{+}}^{I=0}$ as three examples to test different choices of the basis wave functions. The results are presented in ~\ref{tab:comGEMbasis}. It should be noticed that the basis wave functions which yield a lower ground state are more precise according to the variational principle. Our numerical results show that using different discrete basis functions makes little difference once they are complete. Meanwhile, including both the dimeson and diquark-antidiquark spatial functions is very important. Otherwise, one can obtain biased results.  For example, neglecting the dimeson spatial wave functions makes the bounded $[bc\bar{n}\bar{n}]_{J^{P}=2^{+}}^{I=0}$ states unbound in the result of $\Psi_E$. The absence of the diquark-antidiquark spatial wave functions in $\Psi_D$ makes the deeply bound $[cc\bar{n}\bar{n}]_{J^{P}=1^{+}}^{I=0}$ state in SLM model a loosely bound state. In our final results, we choose the most general $\Psi_A$ as our basis functions of GEM.

\begin{table*}[htp]
    \centering
        \caption{Comparisons of the GEM results in different basis functions in Eqs.~\eqref{eq:GEMwave}. The energies are with respect to the lowest relevant thresholds (in units of MeV). ``NB" represents that there is no bound solution. ``$\hookrightarrow$Excited" labels the excited bound state solutions. }
    \label{tab:comGEMbasis}
\begin{tabular}{ccccccccccccccccc}
\hline \hline
 \multirow{2}{*}{Systems} & \multirow{2}{*}{Thresh.}& \multicolumn{5}{c}{AL1-GEM} & \multicolumn{5}{c}{AP1-GEM} & \multicolumn{5}{c}{SLM-GEM}\tabularnewline
  \cmidrule(r){3-7}\cmidrule(r){8-12}\cmidrule(r){13-17}
 &  & A & B & C & D & E & A & B & C & D & E & A & B & C & D & E\tabularnewline\hline\hline

$[cc\bar{n}\bar{n}]_{J^{P}=1^{+}}^{I=0}$ & $DD^{*}$ & -14.0 & -13.6 & -14.0 & NB & NB & -22.2 & -21.7 & -22.2 & -0.94 & NB & -189.6 & -188.9 & -189.6 & -5.9 & -139.1\tabularnewline
$[bb\bar{n}\bar{n}]_{J^{P}=1^{+}}^{I=0}$ & $\bar{B}\bar{B}^{*}$ & -151.6 & -151.2 & -151.6 & -78.5 & -129.5 & -174.8 & -174.5 & -174.8 & -102.8 & -154.7 & -359.6 & -359.4 & -359.6 & -103.1 & -336.5\tabularnewline
$\hookrightarrow$Excited &  & -0.70  & -0.46 & -0.70 &  &  & -3.3 & -3.0 & -3.3 &  &  & -66.0 & -65.5 & -66.0 &  & \tabularnewline

$[bc\bar{n}\bar{n}]_{J^{P}=2^{+}}^{I=0}$ & $D^{*}\bar{B}^{*}$ & -2.9 & -2.7 & -2.9 & -2.3 & NB & -4.4 & -4.2 & -4.4 & -3.9 & NB & -2.4 & -2.2 & -2.4 & -1.8 & NB\tabularnewline
\hline 
\hline
\end{tabular}
\end{table*}

\subsection{Resonating group method}
\begin{figure}[htp]
	\centering  \includegraphics[width=0.3\textwidth]{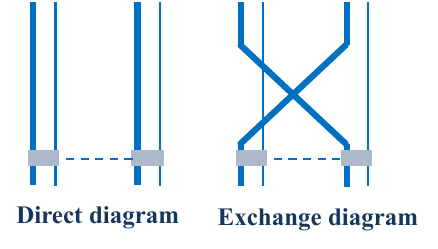}
	\caption{The direct and exchange diagrams in RGM.}\label{fig:RGM}
\end{figure}

In this work, we choose the formalism of RGM in momentum space~\cite{Entem:2000mq}. In the RGM formalism, the  wave functions of the tetraquark states are formulated as,
\begin{equation}
\Psi(\bm{P},\bm{p}_{1},\bm{p}_{2})=\mathcal{A}[\psi_{M1}(\bm{p}_{1})\psi_{M2}(\bm{p}_{2})\psi_{12}(\bm{P})\chi_{csf}^{M1M2}]\,,~\label{eq:RGMwv}
\end{equation}
where $\psi_{M1}$ and $\psi_{M2}$ are spatial wave functions of two mesons, with the $\bm{p}_1$
 and $\bm{p}_2$ the relative momentum of the quark and antiquark inside two mesons, respectively. Accordingly, the discrete wave functions $\chi_{csf}^{M1M2}$ are also the dimeson type ones. In our calculation, we obtain the meson wave functions from the GEM.  The $\psi_{12}$ is the relative wave functions of the two mesons with the corresponding relative momentum $\bm{P}$. We could act $\psi_{M1}\psi_{M2}\chi_{csf}^{M1M2}$  on the Schrodinger equation from the left, $\hat{H}\Psi(\bm{P},\bm{p}_{1},\bm{p}_{2})=E\Psi(\bm{P},\bm{p}_{1},\bm{p}_{2})$. We obtain an equation of $\psi_{12}$,
 \begin{eqnarray}
\int d^{3}\bm{P}\left[V_{D}(\bm{P}',\bm{P})+K_{Ex}(\bm{P}',\bm{P})\right]\psi_{12}(\bm{P}) \nonumber\\
+\left(\frac{\bm{P}'^{2}}{2\mu_{M1M2}}-E\right)\psi_{12}(\bm{P}')	=0 , ~\label{eq:RGM}
 \end{eqnarray}
 where $V_D$ and $K_{Ex}$ are the kernels stemming from the direct diagrams and exchange diagrams in Fig.~\ref{fig:RGM}. $\mu_{M1M2}$ is the reduced mass of the two clusters. In our calculation, the explicit $V_D$ and $K_{Ex}$ can be derived from the meson wave functions. We can solve Eq.~\eqref{eq:RGM} by performing the partial wave expansion and dicretizing the magnitude of the $\bm{P}$ and $\bm{P}'$. In our calculation, we choose the coupled-channel formalism and sum over all the possible two ground meson states with the S-wave relative angular momentum in the wave function of Eq.~\eqref{eq:RGMwv}. The Eq.~\eqref{eq:RGM} becomes the coupled-channel integral equations accordingly. One can find details about RGM in momentum space in Ref.~\cite{Entem:2000mq}.

It should be noticed that the diquark-antidiquark spatial wave functions are absent in the trail functions of RGM. If we use the notations of Sec.~\ref{subsec:GEM}, the wave functions is similar to
 \begin{equation}
     \Psi={\cal A}\left(\chi_{f}\otimes\chi_{cs}^{\text{dimeson}}\otimes\psi^{\text{\text{dimeson}}}\right).\label{eq:wvdimeson}
 \end{equation}
The trial functions are not as general as $\Psi_A$ used in GEM. Meanwhile, the meson wave functions are determined which correspond to the free mesons. The distortion effect of the meson wave functions within the tetraquark bound states is also neglected.  Thus, from the variational principle, we expect the RGM will give a higher solutions than those from GEM.

\subsection{Diffusion Monte Carlo method}
Unlike the two previous methods based on the basis expansion (essentially the variational method), the DMC is a kind of projection Monte Carlo method. One can find the detailed formalism in Refs.~\cite{Ma:2022vqf,Ma:2023int}. To make this paper self-contained, we introduce it briefly. The imaginary time Sch\"ordinger equation reads,
\begin{equation}
    -\frac{\partial\Psi(\boldsymbol{R},t)}{\partial t}	=\left[\frac{1}{2\mu}\boldsymbol{\nabla}^{2}+V(\boldsymbol{R})-E_{R}\right]\Psi(\boldsymbol{R},t),
\end{equation}
with
\begin{align}
    \Psi(\boldsymbol{R},t)	=\sum_{i}c_{i}\Phi_{i}(\boldsymbol{R})e^{-[E_{i}-E_{R}]t},
\end{align}
where $E_R$ is a shift parameter of the energy. $\Phi_i$ are the eigenstates with the eigenvalue $E_i$. One can see if we take the $E_R$ to approach the ground state energy, all $\Phi_i$ except the $\Phi_0$ will be suppressed exponentially after a long-time evolution. 

\begin{figure}[htp]
	\centering  \includegraphics[width=0.4\textwidth]{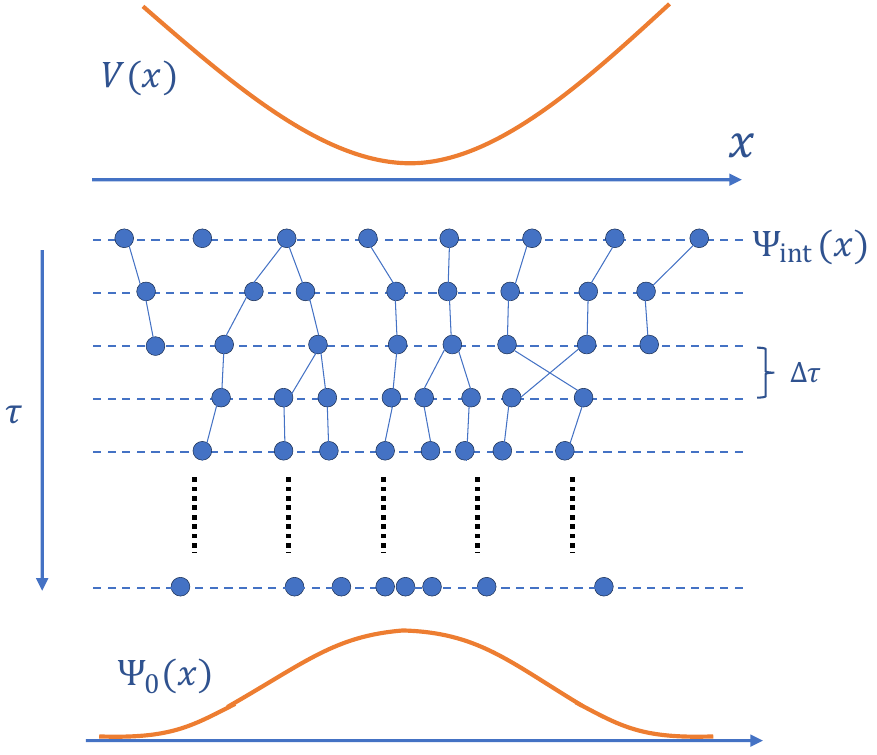}
	\caption{Illustration of the DMC method~\cite{kent1999techniques}, where the wave function is sampled by the walkers.}\label{fig:DMC}
\end{figure}

The DMC is implemented by sampling the wave function with walkers. The distribution of walkers represents the wave function.  The imaginary time Schr\"odinger equation is actually a diffusion equation with the source and sink. As shown in Fig~\ref{fig:DMC}, one can start from a $\Psi_{int}$ which is not orthogonal to the ground state wave function. For every small time step, the walkers will perform a random walk (diffusion process) and experience the death or birth (branch process). In the branch process, one walker could be replicated for several times or be deleted. After a long-time evolution, the distribution of the walkers will approach the ground state wave function. For a practical calculation, the importance sampling is adopted where in addition to the diffusion and branch process, there is an extra drift process (see~\cite{Ma:2022vqf,Ma:2023int} for details).  

As shown in Sec.~\ref{subsec:GEM}, one needs to use either very general or very proper trial functions in the variational method-based approaches to get accurate solutions of a few-body problem. In other words, one should assign a priori clustering behavior to solve the tetraquark states efficiently.  However, DMC could get the ground state energy  without the pre-assignment of the clustering behaviors. In principle, the wave function space allowed by the DMC calculation could be very general. The correct cluster behaviors could be obtained automatically after a long-time evolution. It has been proved that in molecular physics~\cite{suhm1991quantum}, solid physics~\cite{Foulkes:2001zz}, and nuclear physics~\cite{Carlson:2014vla}, the DMC method is very efficient and precise. In hadronic physics, the DMC method has been used in quark models in several works~\cite{Bai:2016int,Gordillo:2020sgc,Gordillo:2021bra,Alcaraz-Pelegrina:2022fsi,Gordillo:2022nnj,Ma:2022vqf,Gordillo:2023tnz}. However, some advantages of the DMC have not been realized in the past calculations. For example, in Ref.~\cite{Gordillo:2020sgc}, the authors performed the calculations for the fully tetraquark states via DMC and obtained many solutions above the corresponding the dimeson thresholds. In principle, one should get the dimeson thresholds if there are no bound state solutions. 

Compared with the electron systems in the molecular physics and solid physics, and the nucleon systems in the nuclear physics, there are two distinct features in the multiquark systems. First, there are complicated color structures for the multiquark systems, which result in complicated discrete wave functions or more coupled channels. Meanwhile, the confinement effect makes the dimeson thresholds the only meaningful thresholds. The four quark threshold and the triquark-quark thresholds are meaningless. To suit the calculations of the multiquark systems, one has to adjust the implements of the DMC method. 

In Ref.~\cite{Ma:2022vqf}, we improved the implements of the DMC and finally got the dimeson thresholds for the fully tetraquark systems by taking more coupled channels into considerations. In Ref.~\cite{Ma:2023int}, we adopted the same strategies to calculate the possible tetraquark bound states of the doubly heavy tetraquark states. In this work, we will further compare the results from the DMC with those from the variatonal methods, GEM and RGM.

\section{Numerical results}~\label{sec:num}

We adopt the GEM and RGM to calculate the fully, triply, doubly and singly heavy tetraquark states. We also compare the results from GEM and RGM with those from the DMC for the doubly heavy tetraquark states.  In our tetraquark calculations, we focus on the difference between the tetraquark masses and the lowest dimeson thresholds in the same quark potential models rather than the absolute values. In Figs.~\ref{fig:QQ}, \ref{fig:QQom} and  \ref{fig:Qs}, we present the possible bound solutions, where the theoretical results are shifted to align the relevant dimeson thresholds to the experimental ones.

\subsection{Fully heavy and triply heavy tetraquark}
We calculate the $J^P=0^+,1^+,2^+$ fully heavy tetraquark states with the following quark contents, 
\begin{align}
\Biggl\{\begin{array}{lll}
[cc\bar{c}\bar{c}]^{C=\pm}\,, & [bb\bar{b}\bar{b}]^{C=\pm}\,, & bb\bar{b}\bar{c}\,,\\{}
[bc\bar{b}\bar{c}]^{C=\pm}\,, & bb\bar{c}\bar{c}\,,         & cc\bar{c}\bar{b}\,,
\end{array}
\end{align}
where $C=\pm$ represents that both states with even and odd C-parities are investigated. The three quark models with different few-body methods yield consistent results. There do not exist the bound state solutions for these systems. These results agree with those qualitatively in Refs.~\cite{Wang:2019rdo,Liu:2019zuc,An:2022qpt,Wang:2022yes,Ortega:2023pmr} where the AL1 and SLM were used, respectively. In lattice QCD simulations, it was shown that there are no bound states of $J^{PC}=0^{++}$, $1^{+-}$ and $2^{++}$ $bb\bar{b}\bar{b}$ bound states below the non-interacting dimeson thresholds~\cite{Hughes:2017xie}. In Ref.~\cite{Junnarkar:2018twb}, the lattice QCD simulations disfavor the existence of the stable spin-0 $bb\bar{c}\bar{c}$ state. These lattice QCD results are also consistent with our findings.

For the triply heavy tetraquark states, we perform the calculations to find possible bound states with $J^P=0^+,1^+, 2^+$ for the following systems,
\begin{align}
        \Biggl\{\begin{array}{llll}
bb\bar{b}\bar{n}\,, & bb\bar{b}\bar{s}\,, & bb\bar{c}\bar{n}\,, & bb\bar{c}\bar{s}\,,\\
cc\bar{c}\bar{n}\,, & cc\bar{c}\bar{s}\,, & cc\bar{b}\bar{n}\,, & cc\bar{b}\bar{s}\,,\\
cb\bar{b}\bar{n}\,, & cb\bar{b}\bar{s}\,, & cb\bar{c}\bar{n}\,, & cb\bar{c}\bar{s}\,.
\end{array}
\end{align}
Our results indicate that there is no bound solution below the relevant dimeson thresholds, which is consistent with results in Ref.~\cite{Lu:2021kut}. The $bb\bar{c}\bar{q}$ tetraquarks were also investigated in lattice QCD~\cite{Junnarkar:2018twb,Hudspith:2020tdf}, where the existence of the bound solutions is inconclusive. It was shown that there is a spin-1 $bb\bar{c}\bar{s}$ state below the threshold in Ref.~\cite{Junnarkar:2018twb}, where the finite volume effect was not considered. To pin down its existence, more data is needed to handle the finite volume effect.

\subsection{Doubly heavy tetraquark states}

\begin{figure*}[htp]
	\centering 
   \includegraphics[width=1\textwidth]{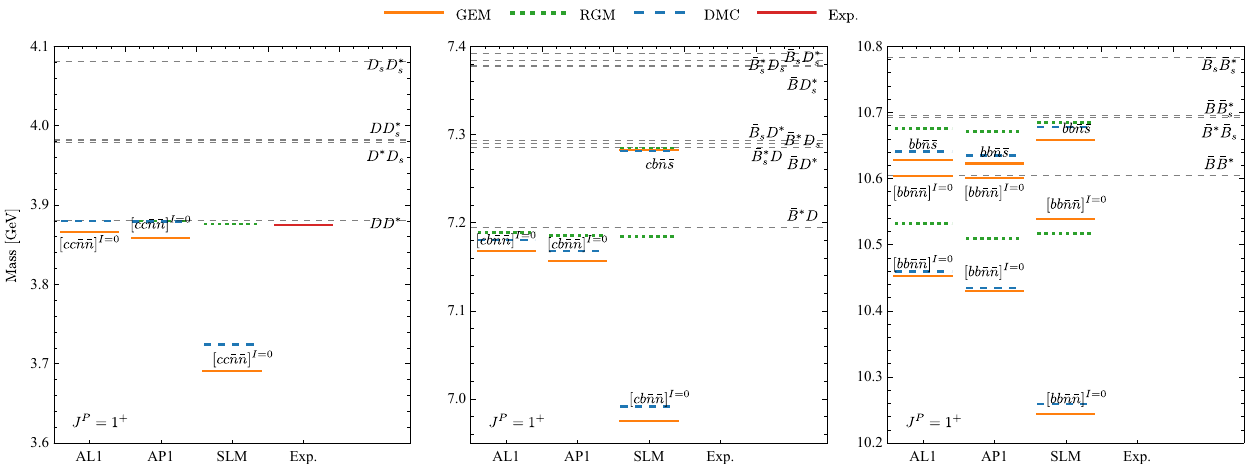}
   \includegraphics[width=1\textwidth]{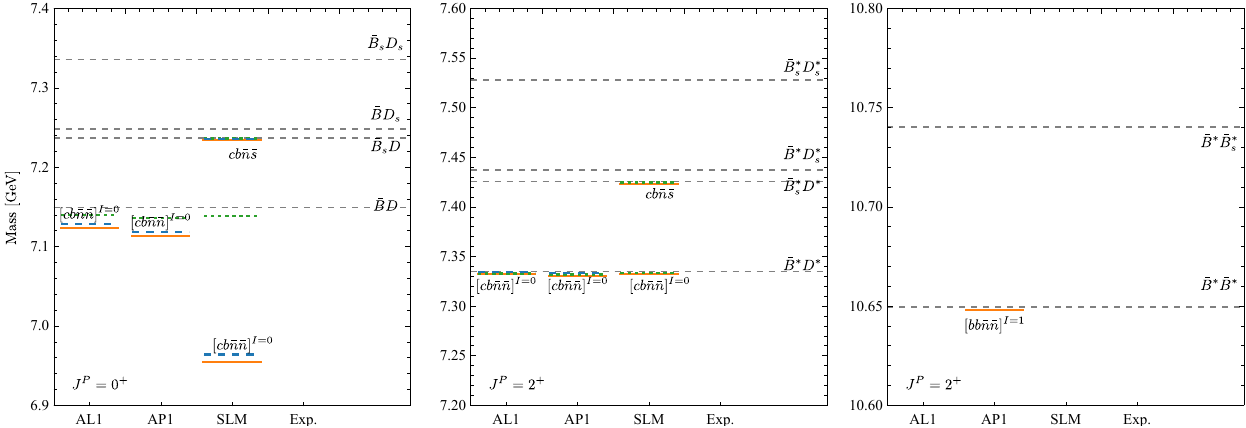}
	\caption{Bound states of the $QQ\bar{q}\bar{q}$ systems from GEM, RGM and DMC in three quark models. The relevant theoretical dimeson thresholds are aligned to the physical ones.}\label{fig:QQ}
\end{figure*}

We investigate the $J^P=0^+,1^+$ and $2^+$ doubly heavy tetraquark states with the following quark contents,
\begin{align}
    \Biggl\{\begin{array}{ccc}
[bb\bar{n}\bar{n}]^{I=0,1}\,, & bb\bar{n}\bar{s}\,, & bb\bar{s}\bar{s}\,,\\{}
[cc\bar{n}\bar{n}]^{I=0,1}\,, & cc\bar{n}\bar{s}\,, & cc\bar{s}\bar{s}\,,\\{}
[bc\bar{n}\bar{n}]^{I=0,1}\,, & bc\bar{n}\bar{s}\,, & bc\bar{s}\bar{s}\,.
\end{array}
\end{align}
Among them, we get several bound states, which are presented in Fig.~\ref{fig:QQ} and Table~\ref{tab:QQ}. Evidently, the results obtained from different models using various few-body methods exhibit significant divergences.

\begin{table*}
    \centering
    \caption{Bound states of the $QQ\bar{q}\bar{q}$ systems. The energies are with respect to the lowest relevant thresholds (in units of MeV). ``NB" represents that there is no bound solution. ``..." labels systems that are not investigated in the literature. ``$\hookrightarrow$Excited" labels the excited bound state solutions.}\label{tab:QQ}
     \begin{tabular*}{\hsize}{@{}@{\extracolsep{\fill}}cllcccccccccccc@{}}
\hline \hline 
\multirow{2}{*}{$J^{P}$} & \multirow{2}{*}{Systems} & \multirow{2}{*}{Thresh.} & \multicolumn{4}{c}{AL1} & \multicolumn{4}{c}{AP1} & \multicolumn{4}{c}{SLM}\tabularnewline
 \cmidrule(r){4-7} \cmidrule(r){8-11}  \cmidrule(r){12-15} 
 &  &  & GEM & RGM & DMC & \cite{Semay:1994ht} & GEM & RGM & DMC & \cite{Semay:1994ht} & GEM & RGM & DMC & \cite{Ortega:2022efc}\tabularnewline
\hline 
 $0^{+}$ & $[bb\bar{n}\bar{n}]^{I=1}$ & $\bar{B}\bar{B}$ & NB & NB & NB & ... & NB & NB & NB & ... & NB & NB & NB & -13.1\tabularnewline
 & $[bc\bar{n}\bar{n}]^{I=0}$ & $D\text{\ensuremath{\bar{B}}}$ & -26.0 & -9.1 & -21 & 1 & -35.8 & -13.1 & -31 & -13 & -194.9 & -10.7 & -185 & ...\tabularnewline
 & $bc\bar{n}\bar{s}$ & $D\text{\ensuremath{\bar{B}_{s}}}$ & NB & NB & NB & ... &  &  &  & ... & -2.9 & -0.8 & -1 & ...\tabularnewline

 \hline
$1^{+}$ & $[cc\bar{n}\bar{n}]^{I=0}$ & $DD^{*}$ & -14.0 & 1.2 & 0 & 11 & -22.2 & -0.5 & -1 & -1 & -189.6 & -4.2 & -156 & -0.387\tabularnewline

 & $[bb\bar{n}\bar{n}]^{I=0}$ & $\bar{B}\bar{B}^{*}$ & -151.6 & -71.9 & -145 & -142 & -174.8 & -95.3 & -170 & -167 & -359.6 & -87.4 & -345 & -21.9\tabularnewline
 & $\hookrightarrow$Excited  &  & -0.70  &  &  &  & -3.3 &  &  &  & -66.0 &  &  & \tabularnewline
 & $[bb\bar{n}\bar{n}]^{I=1}$ & $\bar{B}\bar{B}^{*}$ & NB & NB & NB & ... & NB & NB & NB & ... & NB & NB & NB & -10.5\tabularnewline
 & $bb\bar{n}\bar{s}$ & $\bar{B}_{s}\bar{B}^{*}$ & -63.8 & -16.5 & -51 & -56 & -69.1 & -20.5 & -57 & -61 & -33.2 & -7.5 & -14 & ...\tabularnewline
 & $[bc\bar{n}\bar{n}]^{I=0}$ & $D\bar{B}^{*}$ & -26.5 & -6.0 & -14 & -5 & -37.8 & -9.2 & -27 & -20 & -219.5 & -10.5 & -203 & ...\tabularnewline
 & $bc\bar{n}\bar{s}$ & $D\bar{B}_{s}^{*}$ & NB & NB & NB & ... & NB & NB & NB & ... & -3.3 & -0.9 & -4 & ...\tabularnewline
 \hline
$2^{+}$ & $[bb\bar{n}\bar{n}]^{I=1}$ & $\bar{B}^{*}\bar{B}^{*}$ & NB & NB & NB & 24 & -1.4 & NB & NB & 24 & NB & NB & NB & -7.1\tabularnewline
 & $[bc\bar{n}\bar{n}]^{I=0}$ & $D^{*}\bar{B}^{*}$ & -2.9 & -2.4 & -1 & ... & -4.4 & -3.3 & -2 & ... & -2.4 & -1.6 & NB & ...\tabularnewline
 & $bc\bar{n}\bar{s}$ & $D^{*}\bar{B}_{s}^{*}$ & NB & NB & NB & ... & NB & NB & NB & ... & -2.4 & -1.4 & NB & ...\tabularnewline
\hline \hline 
\end{tabular*}
\end{table*}

\begin{figure*}[htp]
	\centering 
   \includegraphics[width=1\textwidth]{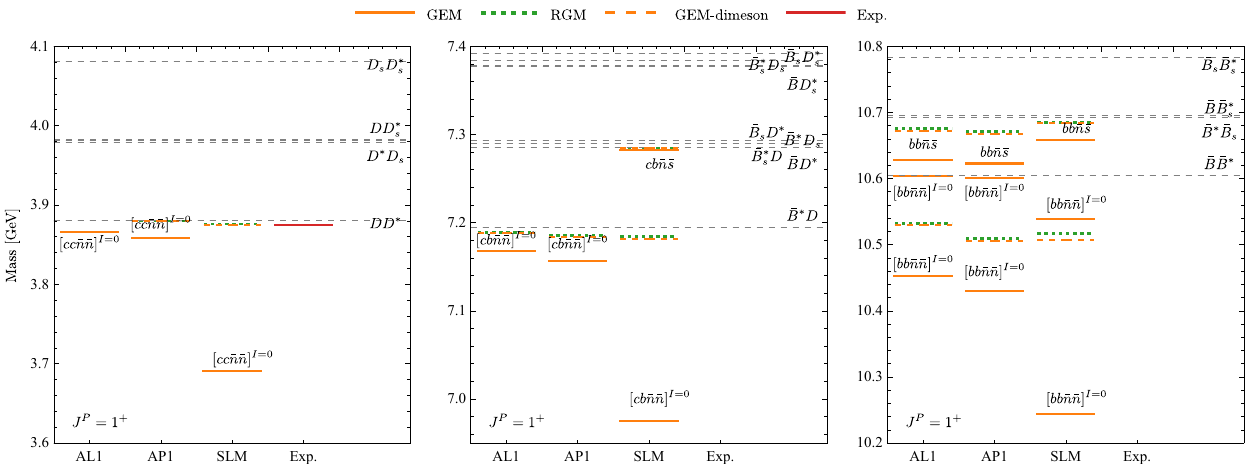}
	\caption{Comparisons of the results from GEM with the general basis function $\Psi_A$ in Eq.~\eqref{eq:GEMwave}, GEM with only the dimeson-type wave function in Eq.~\eqref{eq:wvdimeson} and RGM. The relevant theoretical dimeson thresholds are aligned to the physical ones.}\label{fig:QQom}
\end{figure*}

We first compare two methods based on variational methods, GEM and RGM. As we expected, the GEM gives the lower bound states than the RGM.  For example, for the $J^P=1^+$ $[cc\bar{n}\bar{n}]^{I=0}$ state (the candidate of the experiment $\Tcc$ state), the GEM yields a bound state solution in the AL1 model while the RGM calculation indicates no bound sate. For the same state in the SLM model, GEM gives a very deep bound state with the binding energy about 200 MeV while the binding energy in RGM is about 4 MeV. These disparities arise from the fact that the trial functions or basis functions of GEM are more general than those of RGM where only the dimeson-type wave functions are used. In order to testify the above statement, we also employ the GEM with only the dimeson-type functions, see~\eqref{eq:wvdimeson}, for the $J^P=1^+$ states. We compare the results with those from the RGM and GEM with general wave functions in Fig.~\ref{fig:QQom}. One can see the GEM results with only the dimeson-type trial functions are very similar to those from the RGM. For example, the bound solution of $[cc\bar{n}\bar{n}]^{I=0}$ in AL1 model disappears and the result in SLM model becomes a loosely bound state. We can find the GEM with only the dimeson-type wave functions still yields slightly lower solutions than the RGM. This is because in the implement of the RGM, the meson wave functions is constrained to be the same as those of the free mesons. However, in the GEM-dimeson scheme, the meson wave functions are also determined by the variational parameters, where the possible distortion effect of the meson wave functions is included. 

It should be noticed that the constrained trial functions in RGM could change the results qualitatively. One will miss the bound solution of $J^P=1^+$ $[cc\bar{n}\bar{n}]^{I=0}$ state in AL1 model. Meanwhile, one could identify the $J^P=1^+$ $[bb\bar{n}\bar{n}]^{I=0}$ state as the molecular states rather than the compact tetraquark quark states via RGM. Thus, the GEM with the general basis functions is superior than the RGM.

We can also compare the bound state solutions from DMC in Fig~\ref{fig:QQ} and Table~\ref{tab:QQ} with those from GEM and RGM. It is evident that the majority of DMC results exhibit lower energies compared to the RGM outcomes.  Qualitatively, the DMC results are consistent with those obtained through GEM, as exemplified by the presence of deep bound states for the $J^P=1^+$ $[cc\bar{n}\bar{n}]^{I=0}$ configuration in the SLM model. In essence, the DMC method, devoid of any prior constraints on the clustering behavior of the wave functions, naturally provides the roughly proper clustering behavior in the present implementation. A more comprehensive discussion is available in Ref.~\cite{Ma:2023int}. When compared to GEM results, the masses obtained through DMC generally are higher. For instance, in the case of the $J^P=1^+$ $[cc\bar{n}\bar{n}]^{I=0}$ configuration in the AL1 model, DMC only yields the dimeson thresholds, whereas GEM predicts the existence of a bound state with a binding energy of approximately 14 MeV. It should be noted that the statistical uncertainties in the current DMC calculations are estimated to be at the order of 1 MeV~\cite{Ma:2023int}. Consequently, it is reasonable to conclude that the observed differences stem from systemic uncertainties within the DMC method that have yet to be fully comprehended. One plausible explanation could be that the importance functions employed in Ref.~\cite{Ma:2023int} may not be optimized for addressing the coupled-channel complexities inherent to the tetraquark systems. In principle, the antisymmetrization of the identical Fermions and the associated sign problem should be addressed through proper choices of the importance functions, thereby opening up avenues for future improvements.

Based the above comparisons, it is evident that the GEM is superior than the RGM and DMC method. With this in mind, we will utilize the results obtained through the GEM to conduct a comprehensive comparison across the three distinct quark potential models. 

\subsubsection{$J^P=1^+$}
The experimental $\Tcc$ state is the candidate of $J^P=1^+$ $[cc\bar{n}\bar{n}]^{I=0}$ tetraquark state. Notably, both the PCQM and $\chi$CQM yield the bound state solutions, which can be viewed as the predictions made by these quark models prior to the experimental observation. In the original work of AL1 and AP1 models~\cite{Semay:1994ht}, the existence of the bound states was not conclusively established due to the limitations in computational resources at the time. Meanwhile, the results from AL1 and AP1 are more consistent with the experimental result of $\Tcc$, a very loosely bound state. Conversely, the SLM model suggests the presence of a compact tetraquark state well below the $DD^*$ threshold, with a substantial binding energy of approximately 200 MeV. In Ref.~\cite{Ortega:2022efc}, employing the same SLM, a loosely bound state was achieved through the RGM. The result stems from the constrained basis wave function in RGM. Before the experimental observations, investigations of the $J^P=1^+$ $[cc\bar{n}\bar{n}]^{I=0}$ tetraquark state were undertaken in lattice QCD simulations in Refs. \cite{Cheung:2017tnt,Francis:2018jyb,Junnarkar:2018twb}, but the existence of this state remained inconclusive. Subsequent to the experimental observations, lattice QCD simulations based on L\"uscher's method \cite{Padmanath:2022cvl} and the potential method (HAL QCD method)~\cite{Lyu:2023xro} reported virtual states in this channel.

In all three quark models, we consistently obtain very deep bound state of $J^P=1^+$ $[bb\bar{n}\bar{n}]^{I=0}$, which is the possible heavy quark flavor symmetry partner of $T_{cc}$ state. The three quark models indicate that both the ground states and the first excited states of the system are the bound states. It is worth noticing that the binding energy obtained in SLM is still much larger than those from AL1 and AP1 models. This state has also been extensively investigated in lattice QCD simulations~\cite{Francis:2016hui,Junnarkar:2018twb,Leskovec:2019ioa,Aoki:2023nzp}, which consistently establish the existence of the bound states with the binding energies about 100-200 MeV. Additionally, the static potentials from lattice QCD~\cite{Bicudo:2012qt,Bicudo:2015vta,Bicudo:2016ooe,Bicudo:2021qxj} also indicate the existence of the deeply bound $T_{bb}$ states. Furthermore, these three distinct models also predict the existence of $J^P=1^+$ $[bc\bar{n}\bar{n}]^{I=0}$ bound states. It is worthwhile to notice that the result from SLM is significantly deeper than those from the AP1 and AL1 models. For this state, the lattice QCD simulations have not provided a consistent conclusion~\cite{Francis:2018jyb,Hudspith:2020tdf,Meinel:2022lzo,Padmanath:2023rdu}. 

In our calculations, there is no isovector bound state for the $QQ\bar{n}\bar{n}$ systems. However, in Ref.~\cite{Ortega:2022efc}, a $J^P=1^+$ $[bb\bar{n}\bar{n}]^{I=1}$ bound states were obtained in SLM within the RGM framework. Apparently, this result conflicts with our calculation. \clabel[isovector]{ 
The absence of an isovector bound state in $J^P=1^+$ $[bb\bar{n}\bar{n}]^{I=1}$ system is widely acknowledged and is a consensus reflected in the majority of pertinent publications, such as Ref.~\cite{Yang:2009zzp,Vijande:2009kj,Deng:2018kly}. }

Another state worth attention is the $J=1^+$ $bb\bar{n}\bar{s}$ state. We obtain the bound state below the $\bar{B}_s\bar{B}^*$ threshold with the binding energy 30-70 MeV. In lattice QCD simulations, the existence of this bound state was also implied~\cite{Francis:2016hui,Junnarkar:2018twb,Meinel:2022lzo,Hudspith:2023loy}. 

In addition to the above bound states existing consistently in three different quark models, the SLM model also predicts a $bc\bar{n}\bar{s}$ bound state, which is absent in AL1 and AP1 models.  

\subsubsection{$J^P=0^+$ and $J^P=2^+$}
For the doubly heavy tetraquark states with $J^P=0^+$ and $2^+$, the PCQM and $\chi$CQM both predict the $[bc\bar{n}\bar{n}]^{I=0}$ bound states. The main difference is the $J^P=0^+$ state from SLM is much lower than those from AL1 and AP1. Additionally, the SLM model predicts the extra $bc\bar{n}\bar{s}$ states for $J^P=0^+$ and $2^+$. The AP1 model predicts the extra isovector tensor $bb\bar{n}\bar{n}$ bound state.


\subsection{Singly heavy tetraquark states}

\begin{figure*}[htp]
	\centering 
   \includegraphics[width=1\textwidth]{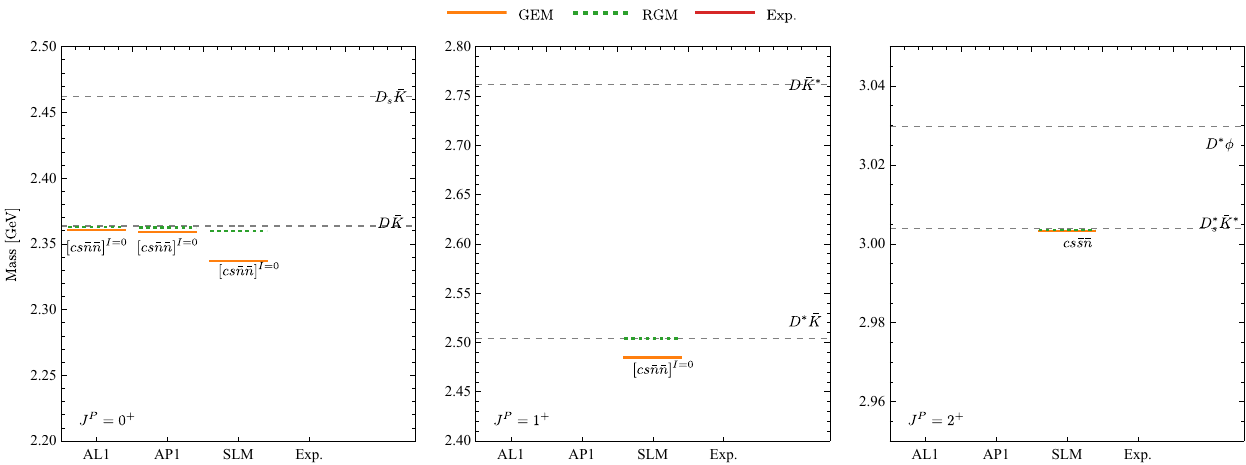}
   \includegraphics[width=1\textwidth]{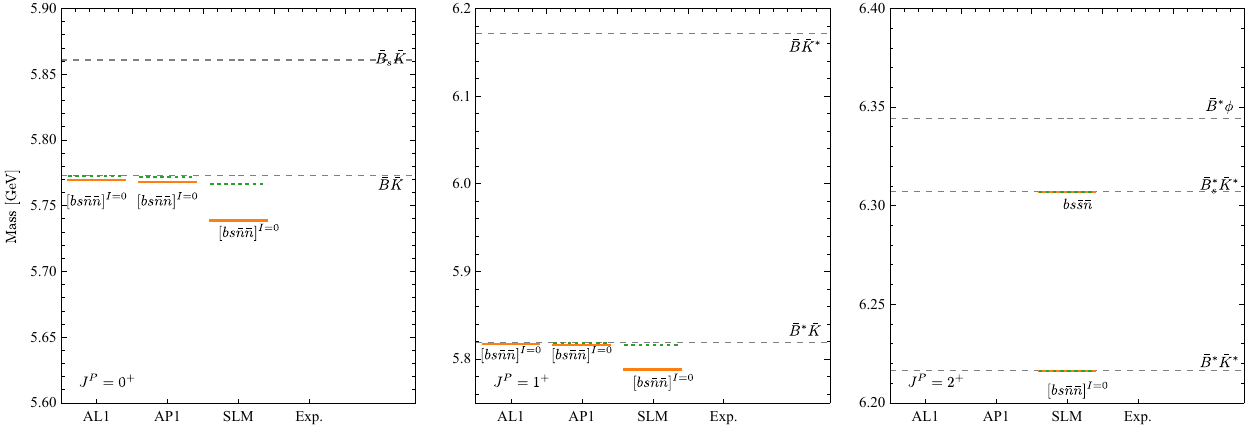}
	\caption{Bound states of the $Qq\bar{q}\bar{q}$ systems from GEM and RGM in three quark models. The relevant theoretical dimeson thresholds are aligned to the physical ones.}\label{fig:Qs}
\end{figure*}

For the singly heavy quark states, we investigate the $J^P=0^+,1^+$ and $2^+$ system with the following quark contents,
\begin{align}
\Biggl\{\begin{array}{llll}
[bn\bar{s}\bar{n}]^{I=1}\,, & [bs\bar{n}\bar{n}]^{I=0,1}\,, & bs\bar{s}\bar{n}\,, & bn\bar{s}\bar{s}\,,\\{}
[cn\bar{s}\bar{n}]^{I=1}\,, & [cs\bar{n}\bar{n}]^{I=0,1}\,, & cs\bar{s}\bar{n}\,, & cn\bar{s}\bar{s}\,,\\{}
[bn\bar{n}\bar{n}]^{I=3/2}\,, & [cn\bar{n}\bar{n}]^{I=3/2}\,, & bs\bar{s}\bar{s}\,, & cs\bar{s}\bar{s}\,,
\end{array}
\end{align}
where the $[Qn\bar{s}\bar{n}]^{I=0}$ and $[Qn\bar{n}\bar{n}]^{I=1/2}$ are not considered because we only focus on the manifestly exotic states. We use the GEM and RGM to solve the spectra in three models. The results are presented in Table~\ref{tab:Qs} and Fig.~\ref{fig:Qs}. 

Comparing the results from GEM and RGM, it is apparent that the results from RGM could be biased due to the constrained basis functions. We still employ the GEM to compare the results from different quark potential models. 

One can see that three quark models all predict the $J^P=0^+$ $[cs\bar{n}\bar{n}]^{I=0}$, $J^P=0^+,1^+$ $[bs\bar{n}\bar{n}]^{I=0}$ bound states. For all the above results, the predictions from SLM tend to be much deeper. It should be noticed that the $J^P=0^+$ $[cs\bar{n}\bar{n}]^{I=0}$ is irrelevant to the experimental $T_{cs0}(2900)$ state which is a resonance state close to the $D^*\bar{K}^*$ threshold. Additionally, the SLM model predicts several extra bound states, $J^P=1^+$ $[cs\bar{n}\bar{n}]^{I=0}$ state, $J^P=2^+$ $cs\bar{s}\bar{n}$ state, $J^P=2^+$ $[bs\bar{n}\bar{n}]^{I=0}$ state, and $J^P=2^+$ $bs\bar{s}\bar{n}$ state.

In Ref.~\cite{Ortega:2023azl}, the authors investigated the spin-0 and spin-1 $cs\bar{n}\bar{n}$ and $cn\bar{s}\bar{n}$ states in SLM and obtained no bound solution, which conflicts with our $J^P=0^+,1^+$ $[cs\bar{n}\bar{n}]^{I=0}$ bound states. In their calculations, several virtual states are found.

\begin{table*}[htp]
    \centering
    \caption{Bound states of the $Qq\bar{q}\bar{q}$ systems. The notations are the same as those in Table~\ref{tab:QQ}.}~\label{tab:Qs}
        \begin{tabular*}{\hsize}{@{}@{\extracolsep{\fill}}cllccccccc@{}}
\hline \hline 
\multirow{2}{*}{$J^{P}$} & \multirow{2}{*}{Systems} & \multirow{2}{*}{Thresh.} & \multicolumn{2}{c}{AL1} & \multicolumn{2}{c}{AP1} & \multicolumn{3}{c}{SLM}\tabularnewline
 \cmidrule(r){4-5} \cmidrule(r){6-7} \cmidrule(r){8-10} 
 &  &  & GEM & RGM & GEM & RGM & GEM & RGM &  \cite{Ortega:2023azl}\tabularnewline
\hline 
$0^{+}$ & $[bs\bar{n}\bar{n}]^{I=0}$ & $\bar{B}\bar{K}$ & -3.4 & -0.7 & -5.1 & -1.2 & -34.4 & -6.6 & ...\tabularnewline
 & $[cs\bar{n}\bar{n}]^{I=0}$ & $D\bar{K}$ & -2.7 & -0.6 & -4.4 & -1.2 & -26.6 & -3.7 & virtual state\tabularnewline
\hline 
$1^{+}$ & $[bs\bar{n}\bar{n}]^{I=0}$ & $\bar{B}^{*}\bar{K}$ & -1.2 & NB & -2.6 & -0.1 & -30.6 & -2.6 & ...\tabularnewline
 & $[cs\bar{n}\bar{n}]^{I=0}$ & $D^{*}\bar{K}$ & NB & NB & NB & NB & -19.4 & -0.3 & NB\tabularnewline
 & $[cs\bar{n}\bar{n}]^{I=1}$ & $D^{*}\bar{K}$ & NB & NB & NB & NB & NB & NB & virtual state\tabularnewline
 \hline 
$2^{+}$ & $bs\bar{s}\bar{n}$ & $\bar{B}_{s}^{*}\bar{K}^{*}$ & NB & NB & NB & NB & -0.4 & -0.2 & ...\tabularnewline
 & $cs\bar{s}\bar{n}$ & $D_{s}^{*}\bar{K}^{*}$ & NB & NB & NB & NB & -0.6 & -0.4 & ...\tabularnewline
 & $[bs\bar{n}\bar{n}]^{I=0}$ & $\bar{B}^{*}\bar{K}^{*}$ & NB & NB & NB & NB & -0.5 & -0.4 & ...\tabularnewline
\hline \hline 
\end{tabular*}

\end{table*}

\section{Discussions and Summary}~\label{sec:sum}

In this study, we conduct benchmark test calculations to investigate the  tetraquark bound states that are manifestly exotic across three distinct quark models (AL1, AP1, and SLM) and employ three different few-body methods (GEM, RGM, and DMC). In the GEM calculations, we find the results is insensitive to the choice of the discrete wave functions once they are complete. However, when it comes to the spatial wave functions, the inclusion of both diquark-antidiquark and dimeson types is imperative for obtaining precise solutions.  We use the GEM with the general wave functions as the standard to compare with the results from the RGM and DMC. 

Our results show that the GEM is superior than RGM and DMC by obtaining the more lower ground state energies and detecting extra bound solutions. Because GEM is a method based on the variational principle, the lower ground state solutions mean more precise results. While RGM can identify the molecular-type states, it tends to produce biased results for the deeply bound states due to its constraints on the basis functions, particularly the dimeson-type wave functions.  DMC can identify both the compact tetraquark states and loosely bound molecular states without a priori assumption about the clustering behaviors of the wave functions. However, when compared to GEM, the current implementation of DMC falls slightly short in terms of precision. This could be attributed to the possibility that the importance functions used in Ref.~\cite{Ma:2023int} may not have been optimized to suit the multiquark systems. Up to now, the advantages of the DMC method, as demonstrated in atomic and molecular physics as well as nuclear physics, has not been fully exploited in the multiquark systems. DMC remains a promising approach distinct from the variational method. The DMC method circumvents the challenges associated with the exponentially growing basis set with number of particles and the intricate integrals tied to few-body forces in the variational approach. Its advantages are likely to become apparent in the case of multiquark states with a large number of quarks~\cite{Alcaraz-Pelegrina:2022fsi} and in systems featuring flux-tube few-body potentials~\cite{Ma:2022vqf}.  It holds potential for further development and refinement.

\clabel[excited]{
In Table~\ref{tab:QQ}, we also consider the potential excited states $[bb\bar{n}\bar{n}]^{I=0}_{J^P=1^+}$. The basis expansion method can be extended to encompass these excited states, although the Ritz theorem in quantum mechanics textbooks typically emphasizes the variational method's role in establishing an upper limit for ground states. The expectation value of the Hamiltonian remains stationary in the vicinity of its discrete eigenvalues. Consequently, with an increase in the dimensions of the basis space, the eigenvalues within this space converge towards the exact solutions of the Hamiltonian, encompassing both ground and excited states. Further details can be found in Ref. \cite{suzuki1998stochastic}. Alternatively, the Diffusion Monte Carlo (DMC) method can be extended to identify excited states as well (see \cite{hammond1994monte}) }

We use the results from GEM to compare the results from three quark models belonging to two types, PCQM (AL1 and AP1) with the minimal OGE interaction and confinement interaction, and $\chi$CQM (SLM) with both two interactions and extra OBE interaction. Our analysis reveals that the SLM models tend to yield more deeper bound states or provide extra bound state solutions than the AL1 and AP1 models. 

We perform the calculations for over 150 tetraquark states for the fully, triply, doubly and singly heavy tetraquark systems with $J^P=0^+,1^+$ and $2^+$. We summarize the bound states existing in all three quark potential models in Table~\ref{tab:sum_bound}. We find there are no fully heavy and triply heavy tetraquark bound states. For the doubly heavy tetraquark systems, we find the $[cc\bar{n}\bar{n}]^{I=0}_{J^P=1^+}$ (candidate state of the experimental $\Tcc$ state), and its heavy quark flavor symmetry partners $[bb\bar{n}\bar{n}]^{I=0}_{J^P=1^+}$ and $[bc\bar{n}\bar{n}]^{I=0}_{J^P=1^+}$ are all bound states. In addition, the $[bc\bar{n}\bar{n}]^{I=0}_{J^P=0^+,2^+}$ and $[bb\bar{n}\bar{s}]_{J^P=1^+}$ are also bound states. It is worthwhile to mention that the $[bb\bar{n}\bar{s}]_{J^P=1^+}$ bound state was also supported by the lattice QCD simulations. For the singly heavy systems, we find the $[bs\bar{n}\bar{n}]^{I=0}_{J^P=0^+,1^+}$ and $[cs\bar{n}\bar{n}]^{I=0}_{J^P=0^+}$ bound states. We hope these stable states against the strong decays may be searched for in the experiments.

We find the SLM quark potential models tend to give more and deeper bound states and the RGM tends to underestimate the bind energy. However, it is still irrational to  discard them. On the one hand, we still have the room to refine the parameters in SLM to fit the experiment results. Meanwhile, the SLM was originally proposed to depict the NN scattering phase~\cite{Entem:2000mq} via RGM with presuming di-baryon clustering behaviors. If one adopts a general trial wave functions of the six quarks, one perhaps obtains quite different solutions of SLM (e.g. deep bound states) instead of the NN scattering states or deuteron. In other words, the SLM model and RGM were used in combination at the birth of this quark potential model. Perhaps, they should still be used in combination. The combination of the SLM and RGM provides a loosely bound solutions for $[cc\bar{n}\bar{n}]^{I=0}_{J=1}$, which is consistent with the experimental $\Tcc$ state. If that is the case, it is still reasonable to use the SLM model when it is assumed in advance that the target state is a molecular state. 

In fact, we do have some hints that the mixing effect between the molecular configurations and the diquark-antidiquark configurations is suppressed. For example, in the flux-tube models, one can model the complicated dynamics of the sea quarks and gluons as the flux-tubes. There could be the dimeson-type flux-tube and the diquark-antidiquark-type (butterfly-type) flux-tube. When considering the possible mixture of the dimeson constructions and diquark-antidiquark constructions, in addition to the valence quark wave functions, one has to consider the flux-tube wave functions, which represent the dynamics of the sea quarks and gluons. The small overlap of the flux-tube wave functions for the different configurations could suppress their mixing effect. In the weak mixing limit, we could get the molecular states without the effect from the diquark-antidiquark configurations. For these states, it is reasonable to use the dimeson basis functions to expand the wave functions like RGM. One can find similar discussion in Ref.~\cite{Wang:2023igz,Andreev:2021eyj,Ma:2023int}. One can also assume the SLM is a quark potential model only working for the molecular configurations. In this way, the combination of the RGM and SLM becomes reasonable.

In our benchmark test calculations, we unveil discrepancies among the various quark potential models and few-body methods on the market. Furthermore, when we compare our results with those in employing the same quark potential models, we continue to encounter inconsistencies.  While some of these inconsistencies may be attributed to limitations in computational precision in earlier years, others remain unexplained. As we are rapidly entering the era of the ``genuine" multiquark states, it becomes increasingly vital to conduct additional benchmark tests of quark model calculations, particularly when involving different research groups.

\begin{table}[htp]
    \centering
        \caption{Final results of the bound states that exist in all three quark potential models.}
    \label{tab:sum_bound}
 \begin{tabular*}{\hsize}{@{}@{\extracolsep{\fill}}cccl@{}}
\hline \hline 
$J^{P}$ & \multicolumn{3}{c}{Bound states}\tabularnewline
\hline
$0^{+}$ & $[bc\bar{n}\bar{n}]^{I=0}$ & $[cs\bar{n}\bar{n}]^{I=0}$ & $[bs\bar{n}\bar{n}]^{I=0}$\tabularnewline
$1^{+}$ & $[cc\bar{n}\bar{n}]^{I=0}$ & $[bc\bar{n}\bar{n}]^{I=0}$ & $bb\bar{n}\bar{s}$\tabularnewline
 & $[bb\bar{n}\bar{n}]^{I=0}$ & $[bs\bar{n}\bar{n}]^{I=0}$ & \tabularnewline
$2^{+}$ & $[bc\bar{n}\bar{n}]^{I=0}$ &  & \tabularnewline
\hline \hline 
\end{tabular*}
\end{table}

\begin{acknowledgements}
L.M. is grateful to the helpful discussions with Eric S. Swanson and Alessandro Giachino.  This project was supported by the National
Natural Science Foundation of China (11975033 and 12070131001). This
project was also funded by the Deutsche Forschungsgemeinschaft (DFG,
German Research Foundation, Project ID 196253076-TRR 110). \end{acknowledgements}

\begin{appendix}
    \section{Eliminate the (nearly) redundant bases}~\label{app:reduce}
\clabel[reduce]{
In the variational method based on basis expansion, the final step involves solving the generalized eigenvalue problem:
\begin{equation}
    \mathbb{H}\bm v=\lambda\mathbb{N} \bm v,\quad\mathbb{H}=\langle i|\hat{H}|j\rangle,\quad\mathbb{N}=\langle i|j\rangle.
\end{equation}
Here, $\mathbb{H}$ and $\mathbb{N}$ represent the Hamiltonian matrix and the overlap matrix, respectively. $\lambda$ and $\bm v$ correspond to the eigenvalue and eigenvector, and $| i \rangle$ and $|j\rangle$ denote the basis states, which may not be orthogonal. In a more general context, basis states may exhibit dependencies, either in a rigorous sense, such as complete basis states becoming linearly dependent after antisymmetrization, or in a less strict sense, where almost ``parallel" basis states are considered linearly dependent, taking into account the machine precision truncation error. The presence of (nearly) dependent bases can lead to ill-conditioned matrices, breaking down algorithms designed to solve the general eigenvalue problem. Therefore, for robust results, it is crucial to eliminate (nearly) redundant bases.}

\change{To this end, various methods can be employed, including the Gram–Schmidt process. In our calculations, we opt for the diagonalization of the overlap matrix $\mathbb{N}$,
\begin{equation}
\left[\begin{array}{c}
\mathbb{T}_{a\times n}\\
\text{\ensuremath{\mathbb{M}_{b\times n}}}
\end{array}\right]\mathbb{N}_{n\times n}\left[\begin{array}{cc}
\mathbb{T}_{n\times a}^{\dagger} & \mathbb{M}_{n\times b}^{\dagger}\end{array}\right]=\left[\begin{array}{cc}
\mathbb{D}_{a\times a} & 0\\
0 & 0
\end{array}\right]
\end{equation}
with $\mathbb{T}\mathbb{N}\mathbb{T}^{\dagger}=\mathbb{D}$, where the transformation matrix are decomposed to  two blocks. The total number of bases is denoted as $n=a+b$, with only $a$ of them being linearly independent. The matrix $\mathbb{D}$ is diagonal, and it is apparent that $\mathbb{N}$ is semi-positive-definite, ensuring that the diagonal elements of $\mathbb{D}$ are all positive. The matrix $\mathbb{N}$ becomes positive-definite if and only if the basis states $|i\rangle$ are linearly independent. To  eliminate the (nearly) redundant basis vectors, we introduce a set of new bases $|\beta \rangle$ with a number of $a$,
\begin{equation}
    |\beta\rangle=\sum_{j=1}^n|j\rangle T_{j\beta}^{\dagger},
\end{equation}
that are orthogonal
\begin{equation}
    \langle\alpha|\beta\rangle=\sum_{i,j=1}^nT_{\alpha i}\langle i|j\rangle T_{j\beta}^{\dagger}=d_{\alpha}\delta^{\alpha\beta},
\end{equation}
where $d_\alpha$ is the elements of matrix $\mathbb{D}$. One can further make the set of bases normalized, $|\tilde{\beta}\rangle=|\beta\rangle\sqrt{d_{\beta}} $. One can get the matrix elements of the Hamiltonian under the orthogonal and normalized bases,
\begin{equation}
\langle\tilde{\alpha}|H|\tilde{\beta}\rangle=\sqrt{\mathbb{D}^{-1}}\mathbb{T}\mathbb{H}\mathbb{T}^{\dagger}\sqrt{\mathbb{D}^{-1}}.
\end{equation}
In practical applications, one can establish a tolerance for the eigenvalues of $\mathbb{N}$ and eliminate basis states associated with very small eigenvalues. This approach effectively removes (nearly) redundant bases, contributing to the robustness of the algorithm.}
    
\end{appendix}

\bibliography{ref}

\end{document}